\documentclass[12pt]{article}
\usepackage{a4wide}
\usepackage{latexsym}
\usepackage{amsmath}
\usepackage{amsfonts}
\usepackage{amscd}
\usepackage{cite}
\usepackage{graphicx}
\usepackage{axodraw}
\usepackage{float}

\usepackage{pslatex}
\usepackage[latin1]{inputenc}
\usepackage[OT2,T1]{fontenc}

\newcommand{\bq}{\begin{eqnarray}}
\newcommand{\eq}{\end{eqnarray}}
\newcommand{\eps}{\varepsilon}

\DeclareSymbolFont{cyrletters}{OT2}{wncyr}{m}{n}
\DeclareMathSymbol{\Sha}{\mathalpha}{cyrletters}{"58}

\begin{document}

\thispagestyle{empty}

\begin{flushright}
  MITP/13-035
\end{flushright}

\vspace{1.5cm}

\begin{center}
  {\Large\bf Construction of an effective Yang-Mills Lagrangian with manifest BCJ duality\\
  }
  \vspace{1cm}
  {\large Mathias Tolotti and Stefan Weinzierl\\
\vspace{2mm}
      {\small \em PRISMA Cluster of Excellence, Institut f{\"u}r Physik, }\\
      {\small \em Johannes Gutenberg-Universit{\"a}t Mainz,}\\
      {\small \em D - 55099 Mainz, Germany}\\
  } 
\end{center}

\vspace{2cm}

\begin{abstract}\noindent
  {
The BCJ decomposition is a highly non-trivial property of gauge theories.
In this paper we systematically construct an effective Lagrangian, whose Feynman rules
automatically produce the BCJ numerators.
The effective Lagrangian contains non-local terms.
The difference between the standard Yang-Mills Lagrangian and the effective Lagrangian simplifies to zero.
   }
\end{abstract}

\vspace*{\fill}

\newpage

\section{Introduction}
\label{sect:intro}

In a recent paper \cite{Bern:2008qj} Bern, Carrasco and Johansson (BCJ) 
conjectured that tree amplitudes in massless gauge theories can 
always be put into a form of a pole expansion, such that the kinematical numerators of this form satisfy
anti-symmetry and Jacobi-like relations, whenever the associated colour factors do.
This conjecture has triggered significant 
research in this direction \cite{Bern:2010ue,Bern:2011ia,Bern:2010yg,Bern:2013yya,
BjerrumBohr:2009rd,BjerrumBohr:2010zs,BjerrumBohr:2010yc,BjerrumBohr:2010hn,BjerrumBohr:2012mg,
Stieberger:2009hq,Mafra:2010jq,Mafra:2011kj,Mafra:2011nv,Mafra:2010pn,
Grassi:2011ky,
Sondergaard:2011iv,
Feng:2010my,Jia:2010nz,Chen:2011jxa,Du:2011js,Du:2011se,Fu:2012uy,Du:2013sha,
Vaman:2010ez,
Boels:2011mn,Boels:2012ew,Boels:2013bi,
Oxburgh:2012zr,Saotome:2012vy,Broedel:2011pd,Broedel:2012rc,Cachazo:2012uq}.

The BCJ conjecture has several important consequences:
First of all, the conjecture leads to additional relations between colour-ordered partial amplitudes.
These relations are called the BCJ-relations and they 
reduce the number of independent colour-ordered partial amplitudes.
The colour decomposition of a $n$-point tree amplitude involves $(n-1)!$ colour-ordered partial amplitudes.
The Kleiss-Kuijf relations \cite{Kleiss:1988ne,DelDuca:1999rs} reduce the number of independent ones to $(n-2)!$.
The BCJ relations reduces this number further down to $(n-3)!$.
The BCJ relations have been proven first with methods from 
string theory \cite{BjerrumBohr:2009rd,BjerrumBohr:2010zs,Stieberger:2009hq,Mafra:2010jq,Mafra:2011kj,Mafra:2011nv}
and later within quantum field theory with the help of on-shell recursion relations \cite{Feng:2010my,Jia:2010nz,Chen:2011jxa}.

Secondly, the BCJ relations can be applied to the integrand of loop amplitudes. In this case they relate non-planar amplitudes
to planar amplitudes. This is advantageous, since planar amplitudes are considered to be ``easier'' than non-planar amplitudes.

Thirdly, the BCJ relations have implication for gravity \cite{Bern:2008qj,Bern:2010ue,Bern:2011ia,Bern:2010yg,BjerrumBohr:2010yc}.
In simple terms, gravity amplitudes can be thought of as Yang-Mills amplitudes, where the structure constants of the Lie algebra
have been replaced by another copy of the BCJ numerators.

Given the existence of the BCJ decomposition it is a natural question to ask how to construct the BCJ decomposition.
It is a non-trivial task to find a systematic and efficient procedure to determine the BCJ numerators.
The construction of the BCJ decomposition is complicated by the fact that 
the BCJ numerators are not unique.
In fact, they can be modified by generalised gauge transformations \cite{Bern:2008qj,Vaman:2010ez}.
It should also be noted, that the BCJ numerators are not required to be local.
Up to now, the construction of the BCJ numerators has been considered mainly 
at the  amplitude level \cite{Monteiro:2011pc,BjerrumBohr:2012mg,Fu:2012uy,Du:2013sha}
or by using the pure spinor formalism of string theory and BRST covariant building blocks \cite{Mafra:2011kj,Mafra:2011nv}.

In this paper we take a different approach and ask, if it is possible to write down an effective Lagrangian,
whose Feynman rules automatically produce the correct BCJ numerators.
A first step in this direction was already done in \cite{Bern:2010ue},  where an effective Lagrangian was given which generates the BCJ numerators
for tree amplitudes up to five particles.
In this paper we present a systematic algorithm for the construction of an effective Lagrangian with the property that this Lagrangian generates
the BCJ numerators for tree amplitudes up to $n$ particles.
We give explicit results for the $n=6$ case.
The algorithm proceeds inductively and adds new vertices with increasing valency to the standard Yang-Mills Lagrangian.
These additional terms are a complicated zero, therefore 
the difference between the standard Yang-Mills Lagrangian and the effective Lagrangian simplifies to zero.
However, these additional terms ensure that the BCJ relations hold.
The effective Lagrangian is not unique, reflecting the fact that the BCJ decomposition is not unique.
We therefore have to make a choice and select one particular form of the additional terms from the set of all possible additional terms.
Since the algorithm proceeds inductively, a choice has to be made at each order $n$. 
We point out that choices made at orders $<n$ will affect the set of possible terms at order $n$.
We mention that our algorithm can also be used to construct the most general allowed effective Lagrangian by parametrising the non-uniqueness
with free variables.

This paper is organised as follows:
In section~\ref{sect:review} we review the basic concepts:
We start from the standard Yang-Mill Lagrangian, review the colour decomposition and discuss Jacobi-like relations.
We then introduce a notation for trees and finally state the BCJ decomposition.
Section~\ref{sect:effective_Lagrangian} contains the main part of this article and is devoted to the effective Lagrangian.
We first introduce the principles of the effective Lagrangian by discussing the example for $n=5$.
We then present a systematic algorithm for the construction of the effective Lagrangian for arbitrary $n$.
Finally, we give explicit results for $n=6$.
Section~\ref{sect:conclusions} contains our conclusions.
In an appendix we collected the more technical parts: Appendix~\ref{appendix:colour_ordered_rules} shows 
how to obtain the colour-ordered Feynman rule from an operator with $n$ fields.
Appendix~\ref{appendix:inequivalent_tree_topologies} presents an algorithm to find all inequivalent tree topologies for Jacobi-like relations.

\section{Review of the basic concepts}
\label{sect:review}

In this section we recall the necessary background.
In sub-section~(\ref{sect:setup}) we start from the standard Lagrange density for Yang-Mills theory.
In sub-section~(\ref{sect:colour_decomp}) we specialise to Born amplitudes and discuss the colour decomposition as well as the Kleiss-Kuijf relations.
Sub-section~(\ref{sect:jacobi}) is devoted to Jacobi-like relations.
The following sub-section~(\ref{sect:tree_notation}) introduces our notation in connection with trees.
With this preparation the BCJ decomposition can be stated in sub-section~(\ref{sect:bcj_decomposition}).

\subsection{The conventional Yang-Mills Lagrangian}
\label{sect:setup}

The conventional Lagrange density for Yang-Mills gauge theory is given by
\bq
\label{lagrangian_ym}
{\cal L}_{\mathrm{YM}} & = & 
 - \frac{1}{4} F^a_{\mu\nu} F^{a\;\mu\nu},
\eq
where the field strength is as usual given by
\bq
 F^a_{\mu\nu} = \partial_\mu A^a_\nu - \partial_\nu A^a_\mu + g f^{abc} A^b_\mu A^c_\nu.
\eq
The generators $T^a$ of the gauge group satisfy
\bq
 \left[ T^a, T^b \right] = i f^{abc} T^c,
 & & 
 \mbox{Tr}\; T^a T^b = \frac{1}{2} \delta^{ab}.
\eq
The second equation defines the normalisation of the generators used in this article.
It is convenient to introduce the Lie-algebra valued field
\bq
 {\bf A}_\mu & = & \frac{g}{i} T^a A^a_\mu,
\eq
together with the corresponding field strengths
\bq
\label{def_field_strength}
 {\bf F}_{\mu\nu} & = & \partial_\mu {\bf A}_\nu - \partial_\nu {\bf A}_\mu + \left[ {\bf A}_\mu, {\bf A}_\nu \right].
\eq
In terms of these quantities the Yang-Mills Lagrangian can be written as
\bq
\label{lagrangian_ym2}
{\cal L}_{\mathrm{YM}} & = & 
 \frac{1}{2g^2} \; \mbox{Tr} \; {\bf F}_{\mu\nu} {\bf F}^{\mu\nu}.
\eq
As usual we need to fix the gauge. In this paper we work in Feynman gauge.
The gauge fixing term is given by
\bq
 {\cal L}_{\mathrm{GF}}
 & = &
 \frac{1}{g^2} \; \mbox{Tr} \; \left( \partial^\mu {\bf A}_\mu \right) \left( \partial^\nu {\bf A}_\nu \right).
\eq
The gauge-fixing procedure will also introduce Faddeev-Popov ghosts.
However, in this paper we are mainly concerned with Born amplitudes to which ghosts do not contribute.

A typical example of a non-Abelian gauge theory is QCD with the gauge group $SU(3)$.
In analogy with QCD we will call in this paper the gauge bosons ``gluons'' and the gauge degrees of freedom ``colour degrees of freedom''.
However, in our paper nothing is specific to the gauge group $SU(3)$ and all arguments are valid for a general gauge group $G$.

\subsection{Colour decomposition}
\label{sect:colour_decomp}

The tree level amplitude with $n$ external gluons may be written in 
the form
\bq
\label{colour_decomposition_1}
{\cal A}_n(1,...,n) & = &
 g^{n-2} 
 \sum\limits_{\sigma \in S_{n}/Z_{n}} 2 \; \mbox{Tr} \left(
 T^{a_{\sigma(1)}} ... T^{a_{\sigma(n)}} \right)
 A_{n}\left( \sigma_1, ..., \sigma_n \right), 
\eq
where the sum runs over all non-cyclic permutations of the external gluon legs.
The quantities $A_n$, called the partial amplitudes, contain the 
kinematic information.
They are colour-ordered, i.e. only diagrams with a particular cyclic ordering of the gluons 
contribute.
The partial amplitudes are conveniently calculated from colour-ordered Feynman rules.
The colour-ordered Feynman rules for the three-gluon and four-gluon vertex are
\bq
\label{colour_ordered_feynman_rules}
\begin{picture}(100,35)(0,55)
\Vertex(50,50){2}
\Gluon(50,50)(50,80){3}{4}
\Gluon(50,50)(76,35){3}{4}
\Gluon(50,50)(24,35){3}{4}
\LongArrow(56,70)(56,80)
\LongArrow(67,47)(76,42)
\LongArrow(33,47)(24,42)
\Text(50,83)[b]{$1$}
\Text(78,35)[lc]{$2$}
\Text(22,35)[rc]{$3$}
\end{picture}
 & = &
 i \left[ 
   \left( p_2^{\mu_1} - p_3^{\mu_1} \right) g^{\mu_2 \mu_3} 
 + \left( p_3^{\mu_2} - p_1^{\mu_2} \right) g^{\mu_3 \mu_1}
 + \left( p_1^{\mu_3} - p_2^{\mu_3} \right) g^{\mu_1 \mu_2}
   \right],
 \nonumber \\
\begin{picture}(100,75)(0,50)
\Vertex(50,50){2}
\Gluon(50,50)(71,71){3}{4}
\Gluon(50,50)(71,29){3}{4}
\Gluon(50,50)(29,29){3}{4}
\Gluon(50,50)(29,71){3}{4}
\Text(72,72)[lb]{$1$}
\Text(72,28)[lt]{$2$}
\Text(28,28)[rt]{$3$}
\Text(28,72)[rb]{$4$}
\end{picture}
 & = &
   i \left[
        2 g^{\mu_1\mu_3} g^{\mu_2\mu_4} - g^{\mu_1\mu_2} g^{\mu_3\mu_4} - g^{\mu_1\mu_4} g^{\mu_2\mu_3}
 \right].
 \nonumber \\
 \nonumber \\
\eq
Due to the cyclic ordering of the partial amplitudes we have
\bq
 A_n\left( 2, 3, ..., n, 1 \right) & = & A_n\left( 1, 2, 3, ..., n \right).
\eq
There are $(n-1)!$ partial amplitudes appearing in eq.~(\ref{colour_decomposition_1}).
We may use the cyclic property to fix leg $1$ in the first position.
The $(n-1)!$ partial amplitudes in eq.~(\ref{colour_decomposition_1}) correspond then to the $(n-1)!$ 
permutations of the legs $2$, ..., $n$.

The $(n-1)!$ partial amplitudes are not independent.
The Kleiss-Kuijf relations \cite{Kleiss:1988ne} give linear relations between these partial amplitudes.
To state the Kleiss-Kuijf relations we let
\bq
 \vec{\alpha} = \left( \alpha_1, ..., \alpha_j \right),
 & & 
 \vec{\beta} = \left( \beta_1, ..., \beta_{n-2-j} \right)
\eq
and $\vec{\beta}^T = ( \beta_{n-2-j}, ..., \beta_1 )$.
The Kleiss-Kuijf relations \cite{Kleiss:1988ne,DelDuca:1999rs} read
\bq
\label{Kleiss_Kuijf}
 A_n\left( 1, \vec{\alpha}, n, \vec{\beta} \right)
 & = & 
 \left( -1 \right)^{n-2-j}
 \sum\limits_{\sigma \in \vec{\alpha} \; \Sha \; \vec{\beta}^T}
 A_n\left( 1, \sigma_1, ..., \sigma_{n-2}, n \right).
\eq
Here, $\vec{\alpha} \; \Sha \; \vec{\beta}^T$ denotes the set of all shuffles of $\vec{\alpha}$ with $\vec{\beta}^T$, i.e.
the set of all permutations of the elements of $\vec{\alpha}$ and $\vec{\beta}^T$, which preserve the relative order of the
elements of $\vec{\alpha}$ and of the elements of $\vec{\beta}^T$.
The Kleiss-Kuijf relations reduce the number of partial amplitudes to $(n-2)!$.
This follows immediately from eq.~(\ref{Kleiss_Kuijf}), which allows to express any partial amplitude, where $n$ does not
appear in the last position as a linear combination of partial amplitudes, where $n$ appears in the last position.
Therefore all partial amplitudes can be expressed in terms of the $(n-2)!$ partial amplitudes
$A_n\left( 1, \sigma_1, ..., \sigma_{n-2}, n \right)$.

We will soon see that due to the BCJ decomposition there are further relations among the partial amplitudes, 
which reduce the number of independent partial amplitudes down to $(n-3)!$.

\subsection{Jacobi-like relations}
\label{sect:jacobi}

The Jacobi identity satisfied by the structure constants of a Lie algebra reads
\bq
  f^{a_1 a_2 b} f^{b a_3 a_4} + f^{a_2 a_3 b} f^{b a_1 a_4} + f^{a_3 a_1 b} f^{b a_2 a_4}
 & = & 0.
\eq
Equivalently the Jacobi identity can be written as
\bq
 \mbox{Tr} \;
 \left( 
        \left[ \left[ T^{a_1}, T^{a_2} \right], T^{a_3} \right]  T^{a_4}
      + \left[ \left[ T^{a_2}, T^{a_3} \right], T^{a_1} \right]  T^{a_4}
      + \left[ \left[ T^{a_3}, T^{a_1} \right], T^{a_2} \right]  T^{a_4}
 \right) 
 & = & 0.
\eq
We may represent the Jacobi relation graphically as follows:
\bq
\label{four_term_relation}
\begin{picture}(100,50)(0,25)
\Vertex(50,30){2}
\Vertex(35,45){2}
\Line(50,5)(50,30)
\Line(50,30)(80,60)
\Line(50,30)(20,60)
\Line(35,45)(50,60)
\Text(20,65)[b]{$1$}
\Text(50,65)[b]{$2$}
\Text(80,65)[b]{$3$}
\Text(50,0)[t]{$4$}
\end{picture}
 +
\begin{picture}(100,50)(0,25)
\Vertex(50,30){2}
\Vertex(35,45){2}
\Line(50,5)(50,30)
\Line(50,30)(80,60)
\Line(50,30)(20,60)
\Line(35,45)(50,60)
\Text(20,65)[b]{$2$}
\Text(50,65)[b]{$3$}
\Text(80,65)[b]{$1$}
\Text(50,0)[t]{$4$}
\end{picture}
 +
\begin{picture}(100,50)(0,25)
\Vertex(50,30){2}
\Vertex(35,45){2}
\Line(50,5)(50,30)
\Line(50,30)(80,60)
\Line(50,30)(20,60)
\Line(35,45)(50,60)
\Text(20,65)[b]{$3$}
\Text(50,65)[b]{$1$}
\Text(80,65)[b]{$2$}
\Text(50,0)[t]{$4$}
\end{picture}
 & = & 0.
 \nonumber \\
 \nonumber \\
\eq
We may take eq.~(\ref{four_term_relation}) as the definition of an abstract Jacobi-like relation.
Of particular interest is the case where
the three-valent vertex is anti-symmetric:
\bq
\label{antisymmetry_relation}
\begin{picture}(60,30)(0,15)
\Vertex(30,20){2}
\Line(30,5)(30,20)
\Line(30,20)(45,35)
\Line(30,20)(15,35)
\Text(15,40)[b]{$1$}
\Text(45,40)[b]{$2$}
\Text(30,0)[t]{$3$}
\end{picture}
 & = &
 -
\begin{picture}(60,30)(0,15)
\Vertex(30,20){2}
\Line(30,5)(30,20)
\Line(30,20)(45,35)
\Line(30,20)(15,35)
\Text(15,40)[b]{$2$}
\Text(45,40)[b]{$1$}
\Text(30,0)[t]{$3$}
\end{picture}
 \nonumber \\
\eq
In this case we may rewrite eq.~(\ref{four_term_relation}) as
\bq
\label{STU_relation}
\begin{picture}(110,30)(0,30)
\Line(10,10)(90,10)
\Vertex(50,30){2}
\Vertex(50,10){2}
\Line(50,10)(50,30)
\Line(50,30)(70,50)
\Line(50,30)(30,50)
\Text(5,10)[r]{$1$}
\Text(30,55)[b]{$2$}
\Text(70,55)[b]{$3$}
\Text(95,10)[l]{$4$}
\end{picture}
 & = & 
\begin{picture}(120,30)(-10,30)
\Line(10,10)(90,10)
\Vertex(35,10){2}
\Vertex(65,10){2}
\Line(35,10)(35,40)
\Line(65,10)(65,40)
\Text(5,10)[r]{$1$}
\Text(35,45)[b]{$2$}
\Text(65,45)[b]{$3$}
\Text(95,10)[l]{$4$}
\end{picture}
-
\begin{picture}(110,30)(-10,30)
\Line(10,10)(90,10)
\Vertex(35,10){2}
\Vertex(65,10){2}
\Line(35,10)(35,40)
\Line(65,10)(65,40)
\Text(5,10)[r]{$1$}
\Text(35,45)[b]{$3$}
\Text(65,45)[b]{$2$}
\Text(95,10)[l]{$4$}
\end{picture}
 \nonumber \\
 \nonumber \\
\eq
Eq.~(\ref{STU_relation}) is usually called a STU-relation.
If eq.~(\ref{STU_relation}) holds we may reduce any tree graph with $n$ external legs and containing only three-valent vertices to
a multi-peripheral form with respect to $1$ and $n$.
We say that a graph is multi-peripheral with respect to $1$ and $n$, if all other external legs connect directly to the line from $1$ to $n$,
i.e. there are no non-trivial sub-trees attached to this line.
A graph in multi-peripheral form can be drawn as
\bq
\begin{picture}(210,60)(0,0)
\Line(10,10)(200,10)
\Vertex(35,10){2}
\Vertex(65,10){2}
\Vertex(95,10){2}
\Vertex(145,10){2}
\Vertex(175,10){2}
\Line(35,10)(35,40)
\Line(65,10)(65,40)
\Line(95,10)(95,40)
\Line(145,10)(145,40)
\Line(175,10)(175,40)
\Text(5,10)[r]{$1$}
\Text(35,45)[b]{$\sigma_2$}
\Text(65,45)[b]{$\sigma_3$}
\Text(120,25)[b]{$...$}
\Text(175,45)[b]{$\sigma_{n-1}$}
\Text(205,10)[l]{$n$}
\end{picture}
\eq
Repeated use of eq.~(\ref{STU_relation}) reduces any tree graph with non-trivial sub-trees attached 
to the line from $1$ to $n$ to a multi-peripheral form.

\subsection{Trees}
\label{sect:tree_notation}

\begin{figure}
\begin{center}
\bq
\begin{picture}(80,40)(0,20)
\Vertex(40,20){2}
\Line(40,5)(40,20)
\Line(40,20)(10,50)
\Line(40,20)(70,50)
\Vertex(30,30){2}
\Line(30,30)(50,50)
\Vertex(20,40){2}
\Line(20,40)(30,50)
\Text(10,55)[b]{$1$}
\Text(30,55)[b]{$2$}
\Text(50,55)[b]{$3$}
\Text(70,55)[b]{$4$}
\Text(40,0)[t]{$5$}
\end{picture}
\begin{picture}(80,40)(0,20)
\Vertex(40,20){2}
\Line(40,5)(40,20)
\Line(40,20)(10,50)
\Line(40,20)(70,50)
\Vertex(30,30){2}
\Line(30,30)(50,50)
\Vertex(40,40){2}
\Line(40,40)(30,50)
\Text(10,55)[b]{$1$}
\Text(30,55)[b]{$2$}
\Text(50,55)[b]{$3$}
\Text(70,55)[b]{$4$}
\Text(40,0)[t]{$5$}
\end{picture}
\begin{picture}(80,40)(0,20)
\Vertex(40,20){2}
\Line(40,5)(40,20)
\Line(40,20)(10,50)
\Line(40,20)(70,50)
\Vertex(20,40){2}
\Line(20,40)(30,50)
\Vertex(60,40){2}
\Line(60,40)(50,50)
\Text(10,55)[b]{$1$}
\Text(30,55)[b]{$2$}
\Text(50,55)[b]{$3$}
\Text(70,55)[b]{$4$}
\Text(40,0)[t]{$5$}
\end{picture}
\begin{picture}(80,40)(0,20)
\Vertex(40,20){2}
\Line(40,5)(40,20)
\Line(40,20)(10,50)
\Line(40,20)(70,50)
\Vertex(50,30){2}
\Line(50,30)(30,50)
\Vertex(40,40){2}
\Line(40,40)(50,50)
\Text(10,55)[b]{$1$}
\Text(30,55)[b]{$2$}
\Text(50,55)[b]{$3$}
\Text(70,55)[b]{$4$}
\Text(40,0)[t]{$5$}
\end{picture}
\begin{picture}(80,40)(0,20)
\Vertex(40,20){2}
\Line(40,5)(40,20)
\Line(40,20)(10,50)
\Line(40,20)(70,50)
\Vertex(50,30){2}
\Line(50,30)(30,50)
\Vertex(60,40){2}
\Line(60,40)(50,50)
\Text(10,55)[b]{$1$}
\Text(30,55)[b]{$2$}
\Text(50,55)[b]{$3$}
\Text(70,55)[b]{$4$}
\Text(40,0)[t]{$5$}
\end{picture}
\nonumber \\ \nonumber
\eq
\caption{\label{tree_collection}
The cyclic-ordered rooted trees with five external legs and only three-valent vertices.
}
\end{center}
\end{figure}
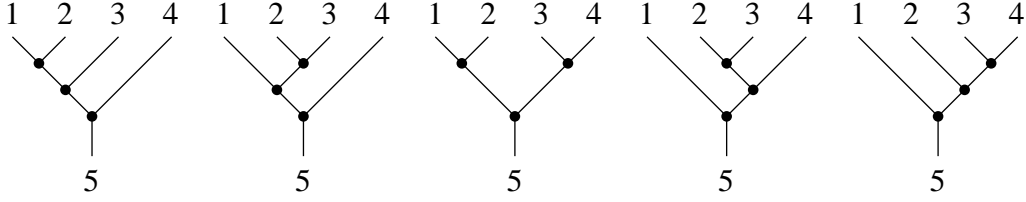

In this paper we are concerned with trees, 
which have a fixed cyclic ordering of the external legs and
which contain only cubic vertices.
Let us assume that we consider trees with $n$ external legs, which are labelled clockwise $(1,2,...,n)$.
If we single out one specific external leg (usually we take the last leg $n$),
we speak of a rooted tree, the root being given by the external leg which we singled out.
We may specify a rooted tree by brackets involving the remaining legs, for example
\bq
 \left[ \left[ 1, 2 \right], 3 \right]
\eq
denotes the rooted tree
\bq
\begin{picture}(100,50)(0,25)
\Vertex(50,30){2}
\Vertex(35,45){2}
\Line(50,5)(50,30)
\Line(50,30)(80,60)
\Line(50,30)(20,60)
\Line(35,45)(50,60)
\Text(20,65)[b]{$1$}
\Text(50,65)[b]{$2$}
\Text(80,65)[b]{$3$}
\Text(50,0)[t]{$4$}
\end{picture}.
 \nonumber \\
 \nonumber \\
\eq
For $n=3$ there is only one rooted tree
\bq
 T^{(3)}_1 & = & \left[ 1, 2 \right],
\eq
for $n=4$ there are two rooted trees
\bq
 T^{(4)}_1 = \left[ \left[ 1, 2 \right], 3 \right],
 & &
 T^{(4)}_2 = \left[ 1, \left[ 2, 3 \right] \right],
\eq
while for $n=5$ there are five trees
\bq
 & &
 T^{(5)}_1 = \left[ \left[ \left[ 1, 2 \right], 3 \right], 4 \right],
 \;\;\;
 T^{(5)}_2 = \left[ \left[ 1, \left[ 2, 3 \right] \right], 4 \right],
 \;\;\;
 T^{(5)}_3 = \left[ \left[ 1, 2 \right], \left[ 3, 4 \right] \right],
 \nonumber \\
 & & 
 T^{(5)}_4 = \left[ 1, \left[ \left[ 2, 3 \right], 4 \right] \right],
 \;\;\;
 T^{(5)}_5 = \left[ 1, \left[ 2, \left[ 3, 4 \right] \right] \right].
\eq
The five trees for $n=5$ are shown in fig.~\ref{tree_collection}.
Multi-peripheral trees with respect to the line from $1$ to $n$ are given in this notation by
\bq
 T^{(n)}_{\mathrm{multi-peripheral}}
 & = &
 \left[\left[\left[ ... \left[\left[\left[ 1, 2 \right], 3 \right], 4 \right], ... \right], n-2 \right], n-1 \right].
\eq
The number $f(n)$ of cyclic-ordered rooted trees with three-valent vertices and $n$ external legs is easily obtained
recursively through
\bq
 f(n) & = & \sum\limits_{i=2}^{n-1} f(i) f(n-i+1),
 \;\;\;\;\;\;
 f(2) = 1.
\eq
A closed formula is given by \cite{Bern:2008qj}
\bq
 f(n) & = & \frac{2^{n-2} (2n-5)!!}{(n-1)!}
 =
 \frac{(2n-4)!}{(n-1)!(n-2)!}.
\eq
We denote by ${\cal T}_n$ the set of all cyclic-ordered rooted trees with three-valent vertices and $n$ 
external legs.

Given a rooted tree $T = [ T_1, T_2 ]$, where $T_1$ and $T_2$ are sub-trees, we define two operators $L$ and $R$, which pick out the left and
the right sub-tree, respectively:
\bq
 L\left(T\right) = T_1,
 & & 
 R\left(T\right) = T_2.
\eq
For a two-tree (or atomic tree) $T=j$ we define
\bq
 L(j) = R(j) = 0.
\eq
In addition we consider non-rooted trees. We define a concatenation operation for two rooted trees:
Let $T_1$ and $T_2$ be two rooted trees with roots $r_1$ and $r_2$.
Then we denote by $(T_1,T_2)$ the non-rooted tree obtained from $T_1$ and $T_2$ by joining the two roots by an edge.
As an example we have:
\bq
\left(
\begin{picture}(60,30)(0,15)
\Vertex(30,20){2}
\Line(30,5)(30,20)
\Line(30,20)(45,35)
\Line(30,20)(15,35)
\Text(15,40)[b]{$1$}
\Text(45,40)[b]{$2$}
\Text(30,0)[t]{$r_{1}$}
\end{picture},
\begin{picture}(60,30)(0,15)
\Vertex(30,20){2}
\Line(30,5)(30,20)
\Line(30,20)(45,35)
\Line(30,20)(15,35)
\Text(15,40)[b]{$3$}
\Text(45,40)[b]{$4$}
\Text(30,0)[t]{$r_{2}$}
\end{picture}
 \right)
 & = &
\begin{picture}(60,30)(-15,15)
\Vertex(15,20){2}
\Vertex(45,20){2}
\Line(15,20)(45,20)
\Line(5,10)(15,20)
\Line(5,30)(15,20)
\Line(55,10)(45,20)
\Line(55,30)(45,20)
\Text(0,10)[r]{$1$}
\Text(0,30)[r]{$2$}
\Text(60,30)[l]{$3$}
\Text(60,10)[l]{$4$}
\end{picture}
 \nonumber \\
\eq
The concatenation operation is symmetric:
\bq
 \left( T_1, T_2 \right) & = & \left( T_2, T_1 \right).
\eq
If $T_1$, $T_2$ and $T_3$ are rooted trees, we have the obvious relations
\bq
 \left( \left[ T_1, T_2 \right], T_3 \right)
 = 
 \left( \left[ T_2, T_3 \right], T_1 \right)
 =
 \left( \left[ T_3, T_1 \right], T_2 \right).
\eq
Every rooted tree can be viewed as a non-rooted tree by simply forgetting that one external leg has been marked as a root.
In order to keep the notation in this paper as simple as possible, we will not distinguish between rooted trees and non-rooted trees
in the case where only properties of non-rooted trees are required.
 
\subsection{The BCJ decomposition}
\label{sect:bcj_decomposition}

We now come back to the Born partial amplitude $A_n(1,...,n)$ with $n$ external gluons.
We label the momenta of the outgoing gluons as $p_1$, ..., $p_n$.
Momentum conservation reads
\bq
 p_1 + p_2 + ... + p_n & = & 0.
\eq
We introduce the notation
\bq
 p_{i..j} & = & p_i + p_{i+1} + ... + p_{j-1} + p_j
\eq
and
\bq
 s_{i...j} & = & p_{i...j}^2.
\eq
We fix leg $n$ as root and we consider a rooted tree
$T \in {\cal T}_{n}$. The rooted tree $T$ has $n$ external legs and $(n-3)$ internal edges.
We denote by $D(T)$ the product of the invariants corresponding to the $(n-3)$ internal edges.
For example
\bq
\label{def_D}
 D\left( \left[\left[\left[ 1, 2 \right], 3 \right], 4 \right] \right)
 & = & 
 s_{1 2} s_{1 2 3},
 \nonumber \\
 D\left( \left[ \left[ 1, 2 \right], \left[ 3, 4 \right] \right] \right)
 & = &
 s_{1 2} s_{3 4}.
\eq
We further denote by $p(T)$ the momentum flowing into the root and by $s(T)=p(T)^2$ the corresponding invariant.
As an example we have
\bq
 p\left(\left[\left[\left[ 1, 2 \right], 3 \right], 4 \right] \right) = p_{1 2 3 4},
 & &
 s\left(\left[\left[\left[ 1, 2 \right], 3 \right], 4 \right] \right) = s_{1 2 3 4}.
\eq
The BCJ decomposition states that the partial amplitudes can be written
as
\bq
\label{bcj_decomposition}
 A_n(1,...,n)
 & = &
 \sum\limits_{T \in {\cal T}_{n}}
 \frac{N(T)}{D(T)},
\eq
where the denominators $D(T)$ have been defined above and where the numerators $N(T)$ satisfy
the following anti-symmetry and Jacobi-like relations:
To state the anti-symmetry relation we consider sub-trees $T_1$, $T_2$, $T_3$ such that
\bq
 T_{12} = \left( \left[ T_1, T_2 \right], T_3 \right),
 \;\;\;\;\;\;
 T_{21} = \left( \left[ T_2, T_1 \right], T_3 \right) 
\eq
are (non-rooted) trees with $n$ external legs. The anti-symmetry relation requires that
\bq
\label{anti_symmetry}
 N\left( T_{12} \right)
 +
 N\left( T_{21} \right)
 & = & 
 0.
\eq
For the Jacobi-like relation we consider sub-trees $T_1'$, $T_2'$, $T_3'$ and $T_4'$ such that
\bq
 T_{123} = \left( \left[ \left[ T_1', T_2' \right], T_3' \right], T_4' \right),
 \;\;\;\;
 T_{231} = \left( \left[ \left[ T_2', T_3' \right], T_1' \right], T_4' \right),
 \;\;\;\;
 T_{312} = \left( \left[ \left[ T_3', T_1' \right], T_2' \right], T_4' \right) 
\eq
are again trees with $n$ external legs. The Jacobi-like relation reads then
\bq
\label{jacobi_relation}
 N\left( T_{123} \right)
+
 N\left( T_{231} \right)
+
 N\left( T_{312} \right)
 & = & 
 0.
\eq
It is a highly non-trivial statement that any partial amplitude $A_n$ can be written in the form of eq.~(\ref{bcj_decomposition}) such that
the numerators satisfy the relations in eq.~(\ref{anti_symmetry}) and eq.~(\ref{jacobi_relation}).
Starting from the conventional Lagrangian in eq.~(\ref{lagrangian_ym}) and using the colour-ordered Feynman rules
of eq.~(\ref{colour_ordered_feynman_rules}) will in general not lead to the above mentioned form.
The problem within the conventional colour-ordered Feynman rules is caused by the four-gluon vertices.
Each four-gluon vertex reduces the number of propagators by one and we have to insert factors like $s/s$ to restore the original
number of propagators.
The challenge is to insert the right factors in the right place.
It is known that the BCJ decomposition is not unique.
One might be tempted to take this as a positive news, since it increases the chances to find a particular BCJ decomposition.
However, this freedom complicates actual calculations: The BCJ decomposition at $n$-points depends on choices made at lower points.
One should think about this freedom as though no additional conditions has been identified which singles out a unique representative.

The fact that partial amplitudes can be written in the form of eq.~(\ref{bcj_decomposition})
with numerators having the properties of eq.~(\ref{anti_symmetry}) and eq.~(\ref{jacobi_relation})
leads to additional relations beyond the Kleiss-Kuijf relations among the partial amplitudes.
These additional relations reduce the number of independent partial amplitudes to $(n-3)!$ \cite{Bern:2008qj}.

At this point, a technical remark on the BCJ decomposition is in order:
Consider a graph contributing to the colour-ordered $n$-point amplitude $A_n$ and consisting only of three-gluon vertices.
This graph contains $(n-2)$ three-valent vertices. Each three-gluon vertex brings in a momentum $q^\mu$, where $q$
is a linear combination of the external momenta. 
For the external particles we have $n$ polarisation vectors $\eps_i^\mu$.
Therefore there are $n$ polarisation vectors $\eps_i^\mu$ and $(n-2)$ momentum vectors $q_j^\mu$, which have to be contracted into
each other.
It follows that there is at least one scalar product $\eps_{i_1} \cdot \eps_{i_2}$, 
where two polarisation vectors are contracted into each other.
Maximally we can have $[n/2]$ scalar products of the type $\eps_{i_1} \cdot \eps_{i_2}$, where $[n/2]$ denotes the largest integer smaller
or equal to $n/2$.
We can therefore write
\bq
\label{decomposition_scalar_products}
 A_n & = & 
 \sum\limits_{j=1}^{[n/2]} A_{n,j},
\eq
where $A_{n,j}$ contains exactly $j$ scalar products of the type $\eps_{i_1} \cdot \eps_{i_2}$.
For the lowest point functions we have
\bq
 A_{4} & = & A_{4,1} + A_{4,2},
 \nonumber \\
 A_{5} & = & A_{5,1} + A_{5,2},
 \nonumber \\
 A_{6} & = & A_{6,1} + A_{6,2} + A_{6,3},
 \nonumber \\
 A_{7} & = & A_{7,1} + A_{7,2} + A_{7,3},
 \nonumber \\
 A_{8} & = & A_{8,1} + A_{8,2} + A_{8,3} + A_{8,4}.
\eq
The leading terms $A_{n,1}$ are obtained from three-gluon vertices only.
The presence of a four-gluon vertex will increase the number of scalar products $\eps_{i_1} \cdot \eps_{i_2}$ by one.
The leading terms $A_{n,1}$ satisfy the BCJ relations automatically.
For the BCJ decomposition we are concerned with the terms $A_{n,j}$ with $j\ge 2$.

\section{The effective Lagrangian}
\label{sect:effective_Lagrangian}

The Yang-Mills Lagrangian can be written as
\bq
\label{YM_vers1}
{\cal L}_{\mathrm{YM}} + {\cal L}_{\mathrm{GF}} & = & 
 \frac{1}{2g^2} \left( {\cal L}^{(2)} + {\cal L}^{(3)} + {\cal L}^{(4)} \right), 
\eq
with
\bq
 {\cal L}^{(2)} & = &
 -2 \mbox{Tr} \; {\bf A}_{\mu} \Box {\bf A}^{\mu},
 \nonumber \\
 {\cal L}^{(3)} & = &
 4 \mbox{Tr} \; \left( \partial_\mu {\bf A}_{\nu} \right) \left[ {\bf A}^{\mu}, {\bf A}^{\nu} \right],
 \nonumber \\
 {\cal L}^{(4)} & = &
 \mbox{Tr} \; \left[ {\bf A}_{\mu}, {\bf A}_{\nu} \right] \left[ {\bf A}^{\mu}, {\bf A}^{\nu} \right].
\eq
We can write the term ${\cal L}^{(4)}$ as
\bq
\label{modify_L4}
 {\cal L}^{(4)} & = &
 - g^{\mu_1\mu_3} g^{\mu_2\mu_4} g_{\nu_1\nu_2}
 \frac{\partial^{\nu_1}_{12} \partial^{\nu_2}_{34} }{\Box_{12}} 
 \; \mbox{Tr} \; \left[ {\bf A}_{\mu_1}, {\bf A}_{\mu_2} \right] \left[ {\bf A}_{\mu_3}, {\bf A}_{\mu_4} \right].
\eq
The subscripts on the derivatives indicate on which fields they act.
The differential operators $\partial^{\nu_1}_{12}$ and $\Box_{12}$ act on the fields ${\bf A}_{\mu_1}$ and ${\bf A}_{\mu_2}$,
but not on the fields ${\bf A}_{\mu_3}$ and ${\bf A}_{\mu_4}$,
while $\partial^{\nu_2}_{34}$ acts on ${\bf A}_{\mu_3}$ and ${\bf A}_{\mu_4}$, but not on ${\bf A}_{\mu_1}$ and ${\bf A}_{\mu_2}$.
A differential operator without any subscripts will act on all fields on its right side.
This notation will be convenient in the sequel.
In eq.~(\ref{modify_L4}) we have introduced a factor $1/\Box_{12}$, which assigns an intermediate propagator to the four-gluon vertex.
Note that this factor cancels against the factor  $-g_{\nu_1\nu_2}\partial^{\nu_1}_{12} \partial^{\nu_2}_{34} = \Box_{12}$.
Leaving the two factors uncanceled keeps the information on the assignment of terms to diagrams with three-valent vertices only.
The trace encodes the colour information, and the propagator corresponds to the tree structure of the colour.
It is therefore convenient to make the following definition: 
We denote by $D^{-1}$ the product of factors $(-1)/\Box$ for all intermediate propagators corresponding to the tree structure of the colour.
The definition is such that in momentum space the inverse operator $D$ agrees with the previously introduced quantity $D$, defined just before
eq.~(\ref{def_D}).
As an example we have:
\bq
 D^{-1} 
 \; \mbox{Tr} \; \left( \left[ \left[ \left[ {\bf A}_{\mu_1}, {\bf A}_{\mu_2} \right], {\bf A}_{\mu_3} \right], {\bf A}_{\mu_4} \right] {\bf A}_{\mu_5} \right)
 & = &
 \frac{\left(-1\right)^2}{\Box_{12} \Box_{123}}
 \; \mbox{Tr} \; \left( \left[ \left[ \left[ {\bf A}_{\mu_1}, {\bf A}_{\mu_2} \right], {\bf A}_{\mu_3} \right], {\bf A}_{\mu_4} \right] {\bf A}_{\mu_5} \right).
 \nonumber
\eq
With this notation we can write ${\cal L}^{(4)}$ as
\bq
\label{modify_L4_v2}
 {\cal L}^{(4)} & = &
 O^{\mu_1\mu_2\mu_3\mu_4}
 D^{-1}
 \; \mbox{Tr} \; \left[ {\bf A}_{\mu_1}, {\bf A}_{\mu_2} \right] \left[ {\bf A}_{\mu_3}, {\bf A}_{\mu_4} \right],
 \nonumber \\
O^{\mu_1\mu_2\mu_3\mu_4} & = &
 g^{\mu_1\mu_3} g^{\mu_2\mu_4} g_{\nu_1\nu_2} \partial^{\nu_1}_{12} \partial^{\nu_2}_{34}
\eq
The interaction term in eq.~(\ref{modify_L4_v2}) leads to the Feynman rule
\bq
\label{four_gluon_vertex_split}
 V^{\mu_1\mu_2\mu_3\mu_4} & = & 
 i \frac{p_{12}^2}{s_{12}} \left( g^{\mu_1\mu_3} g^{\mu_2\mu_4} - g^{\mu_1\mu_4} g^{\mu_2\mu_3} \right)
+
 i \frac{p_{23}^2}{s_{23}} \left( g^{\mu_1\mu_3} g^{\mu_2\mu_4} - g^{\mu_1\mu_2} g^{\mu_3\mu_4} \right).
\eq
Since $p_{12}^2=s_{12}$ and $p_{23}^2=s_{23}$, eq.~(\ref{four_gluon_vertex_split}) reduces to the standard colour-ordered Feynman rule for the four-gluon
vertex given in eq.~(\ref{colour_ordered_feynman_rules}).
However, eq.(\ref{four_gluon_vertex_split}) assigns the individual terms to $s$-channel and $t$-channel diagrams and this assignment generates the 
correct BCJ decomposition up to four gluons.

In order to obtain the BCJ decomposition for higher-point gluon amplitudes we generalise eq.(\ref{YM_vers2}) and write
\bq
\label{YM_vers2}
{\cal L}_{\mathrm{YM}} + {\cal L}_{\mathrm{GF}} & = & 
 \frac{1}{2g^2} \sum\limits_{n=2}^\infty {\cal L}^{(n)}.
\eq
The notation is such that ${\cal L}^{(n)}$ contains $n$ fields.
Of course, in order for eq.~(\ref{YM_vers2}) to agree with eq.~(\ref{YM_vers1}) the terms ${\cal L}^{(n)}$ with $n\ge 5$ 
have to be equivalent to zero.
However, we will write the zero in a special way.
The terms ${\cal L}^{(n)}$ with $n\ge 5$ are constructed such that they generate the BCJ decomposition for gluon amplitudes up to $n$ gluons.
They have a form similar to eq.~(\ref{modify_L4_v2}):
\bq
\label{form_L_n}
 {\cal L}^{(n)} & = &
 \sum\limits_t \sum\limits_{j=2}^{[n/2]}
 O^{\mu_1...\mu_n}_{(n,t,j)}
 \;
 \hat{D}^{-1}
 \;
 \mbox{Tr} \; {\bf T}_{\mu_1...\mu_n}^{(n,t)},
\eq
where $O^{\mu_1...\mu_n}_{(n,t,j)}$ is a differential operator of degree $(n-4)$, 
$\hat{D}^{-1}$ is a pseudo-differential operator of degree $(8-2n)$ to be defined below
and ${\bf T}_{\mu_1...\mu_n}^{(n,t)}$ contains $n$ fields and encodes the colour information.
For a given $n$, there can be more than one term, indicated by the two sums over $t$ and $j$.
The first sum over $t$ allows to have inequivalent trees appearing in ${\bf T}_{\mu_1...\mu_n}^{(n,t)}$,
while the second sum over $j$ corresponds to the number of factors $g^{\mu_{i_1}\mu_{i_2}}$ appearing in each term
of $O^{\mu_1...\mu_n}_{(n,t,j)}$ and corresponds to the decomposition in eq.~(\ref{decomposition_scalar_products}).
As already mentioned we do not need a term for $j=1$.
We require that ${\bf T}_{\mu_1...\mu_n}^{(n,t)}$ vanishes for $n\ge5$ due to the Jacobi identity. This ensures 
${\cal L}^{(n)}=0$ for $n\ge 5$.
To given an example we can take
\bq
\label{example_L_5}
 {\cal L}^{(5)} & = &
  4 g^{\mu_1\mu_3} g^{\mu_2\mu_4} \frac{\partial_{1}^{\mu_5}}{\Box_{123}}
 \left(
 \mbox{Tr} \;  \left[ \left[ \left[ {\bf A}_{\mu_1}, {\bf A}_{\mu_2} \right], {\bf A}_{\mu_3} \right], {\bf A}_{\mu_4} \right] {\bf A}_{\mu_5}
 \right.
 \nonumber \\
 & &
 \left.
  + 
 \mbox{Tr} \; \left[ \left[ {\bf A}_{\mu_3}, {\bf A}_{\mu_4} \right], \left[ {\bf A}_{\mu_1}, {\bf A}_{\mu_2} \right] \right] {\bf A}_{\mu_5}
  + 
 \mbox{Tr} \; \left[ \left[ {\bf A}_{\mu_4}, \left[ {\bf A}_{\mu_1}, {\bf A}_{\mu_2} \right] \right], {\bf A}_{\mu_3} \right] {\bf A}_{\mu_5}
 \right).
 \nonumber
\eq
The term ${\cal L}^{(5)}$ is equal to zero due to the Jacobi identity involving the expressions $[{\bf A}_{\mu_1},{\bf A}_{\mu_2}]$, ${\bf A}_{\mu_3}$ and ${\bf A}_{\mu_4}$.
However, the term ${\cal L}^{(5)}$ generates a five-gluon vertex.
This five-gluon vertex gives a non-vanishing contribution to individual numerators.
In a partial amplitude the sum of all terms related to the five-gluon vertex adds up to zero.
The inclusion of the term ${\cal L}^{(5)}$ in the form above will generate the BCJ decomposition
of the five-gluon amplitude and all lower point amplitudes.

The Jacobi identity ensures that ${\bf T}_{\mu_1...\mu_n}^{(n,t)}$ (and in turn ${\cal L}^{(n)}$) vanishes for $n\ge5$.
It is therefore convenient to introduce for arbitrary Lie algebra-valued expressions ${\bf T}_1$, ${\bf T}_2$ and ${\bf T}_3$ the
notation
\bq
 J\left( {\bf T}_1, {\bf T}_2, {\bf T}_3 \right)
 & = & 
 \left[ \left[ {\bf T}_1, {\bf T}_2 \right], {\bf T}_3 \right]
 +
 \left[ \left[ {\bf T}_2, {\bf T}_3 \right], {\bf T}_1 \right]
 +
 \left[ \left[ {\bf T}_3, {\bf T}_1 \right], {\bf T}_2 \right].
\eq
If ${\bf T}_1$, ${\bf T}_2$ and ${\bf T}_3$ are arbitrary trees, each of the three terms above contains exactly one internal line not contained in the other
two terms. We call this line ``the line marked by the Jacobi identity''.
We can now give the definition of $\hat{D}^{-1}$: When acting on a colour structure containing the Jacobi identity 
the operator $\hat{D}^{-1}$ will give for each term
the previously defined $D^{-1}$ times a factor $(-\Box)$ in the numerator for the marked propagator.
As an example we have
\bq
 \hat{D}^{-1}
  \mbox{Tr} \; J \left( \left[ {\bf A}_{\mu_1}, {\bf A}_{\mu_2} \right], {\bf A}_{\mu_3}, {\bf A}_{\mu_4} \right) {\bf A}_{\mu_5}
 & = &
 - \frac{\Box_{123}}{\Box_{12} \Box_{123}}
 \mbox{Tr} \;  \left[ \left[ \left[ {\bf A}_{\mu_1}, {\bf A}_{\mu_2} \right], {\bf A}_{\mu_3} \right], {\bf A}_{\mu_4} \right] {\bf A}_{\mu_5}
 \nonumber \\
 & &
 -
 \frac{\Box_{34}}{\Box_{12} \Box_{34}}
 \mbox{Tr} \; \left[ \left[ {\bf A}_{\mu_3}, {\bf A}_{\mu_4} \right], \left[ {\bf A}_{\mu_1}, {\bf A}_{\mu_2} \right] \right] {\bf A}_{\mu_5}
 \nonumber \\
 & &
 -
 \frac{\Box_{124}}{\Box_{12} \Box_{124}}
 \mbox{Tr} \; \left[ \left[ {\bf A}_{\mu_4}, \left[ {\bf A}_{\mu_1}, {\bf A}_{\mu_2} \right] \right], {\bf A}_{\mu_3} \right] {\bf A}_{\mu_5}.
\eq
As in the example~(\ref{four_gluon_vertex_split}) we do not cancel common expressions in the numerator and in the denominator.
The denominator defines the relevant tree structure, while the operators in the numerator contribute to the BCJ numerators.
We thus have for the example in eq.~(\ref{example_L_5})
\bq
\label{choice_5}
 {\cal L}^{(5)} & = &
 O^{\mu_1\mu_2\mu_3\mu_4\mu_5}_{(5,1,2)}
 \;
 \hat{D}^{-1}
 {\bf T}_{\mu_1\mu_2\mu_3\mu_4\mu_5}^{(5,1)},
 \nonumber \\
 O^{\mu_1\mu_2\mu_3\mu_4\mu_5}_{(5,1,2)}
 & = &
  -4 g^{\mu_1\mu_3} g^{\mu_2\mu_4} \partial_{1}^{\mu_5},
 \nonumber \\
 \mbox{Tr} \; {\bf T}_{\mu_1\mu_2\mu_3\mu_4\mu_5}^{(5,1)}
 & = &
  \mbox{Tr} \; J \left( \left[ {\bf A}_{\mu_1}, {\bf A}_{\mu_2} \right], {\bf A}_{\mu_3}, {\bf A}_{\mu_4} \right) {\bf A}_{\mu_5}.
\eq
The operator $O^{\mu_1\mu_2\mu_3\mu_4\mu_5}_{(5,1,2)}$ is not unique, for the following three reasons:
First of all the same operator can be written in a different way.
To give an example, the term
\bq
 O^{\mu_1\mu_2\mu_3\mu_4\mu_5}_{(5,1,2),\mathrm{alternative}}
 & = &
  4 g^{\mu_1\mu_5} g^{\mu_2\mu_4} \partial_{2}^{\mu_3},
\eq
will generate exactly the same Feynman rule.
By a suitable relabelling of the indices and the use of momentum conservation (or equivalently the vanishing of a total derivative)
the two operators can be shown to agree.
Secondly, there are operators which generate a Feynman rule, which agrees with the previous one when restricted to the five-particle on-shell
kinematics. In other words, the Feynman rules differ only by terms proportional to $p_i^2$ with $i=1,...,n$.
Thirdly, and most importantly, there are operators which generate a non-vanishing Feynman rule, whose contribution to the BCJ relations vanishes.
An example of this kind would be an operator of the form
\bq
\label{freedom_five}
 4 \lambda \left( g^{\mu_1\mu_3} g^{\mu_2\mu_4} \partial_{1}^{\mu_5} - g^{\mu_1\mu_3} g^{\mu_4\mu_5} \partial_{4}^{\mu_2} \right),
\eq
where $\lambda$ is a free parameter.
This was already noted in \cite{Bern:2010yg}.

\subsection{The algorithm to construct an effective Lagrangian}

We now present an algorithm to construct an effective Lagrangian of the form as in eq.~(\ref{YM_vers2})
\bq
{\cal L}_{\mathrm{YM}} + {\cal L}_{\mathrm{GF}} & = & 
 \frac{1}{2g^2} \sum\limits_{n=2}^\infty {\cal L}^{(n)},
\eq
whose Feynman rules automatically produce the BCJ numerators.
The algorithm proceeds inductively. We assume that all terms ${\cal L}^{(2)}$, ${\cal L}^{(3)}$, ..., ${\cal L}^{(n-1)}$
have already been constructed and we show how to construct ${\cal L}^{(n)}$.
With the remarks of the previous section it is clear that the construction is not unique.
Therefore our algorithm gives one choice out of many possible choices.
We remark that the choices made for the terms ${\cal L}^{(j)}$ with $j<n$ will affect the term ${\cal L}^{(n)}$.
We construct ${\cal L}^{(n)}$ in the form of eq.~(\ref{form_L_n})
\bq
 {\cal L}^{(n)} & = &
 \sum\limits_t \sum\limits_{j=2}^{[n/2]}
 O^{\mu_1...\mu_n}_{(n,t,j)}
 \;
 \hat{D}^{-1}
 \;
 \mbox{Tr} \; {\bf T}_{\mu_1...\mu_n}^{(n,t)}.
\eq
We proceed through the following steps:
\begin{enumerate}
\item We first determine all inequivalent tree topologies ${\bf T}_{\mu_1...\mu_n}^{(n,t)}$ for the Jacobi relations.
An algorithm to do this is given in appendix~\ref{appendix:inequivalent_tree_topologies}.
For a given $n$, this defines a set
\bq
\label{set_topologies}
 \left\{ {\bf T}_{\mu_1...\mu_n}^{(n,1)}, ..., {\bf T}_{\mu_1...\mu_n}^{(n,t_{\mathrm{max}})} \right\}.
\eq
\item For a given $n$, a given $j$ (with $2 \le j \le [n/2]$) and a given $t$ (with $1 \le t \le t_{\mathrm{max}}$) we then consider
all possibilities for the operator $O^{\mu_1...\mu_n}_{(n,t,j)}$.
In general, this operator can be written as a sum of terms, each term containing $j$ factors of the metric tensor.
The operator is of degree $(n-4)$. Therefore it follows that the operator contains $(n-2j)$ derivatives with open indices, while $(2j-4)$ derivatives
are contracted into each other.
We can characterise each term in the operator $O^{\mu_1...\mu_n}_{(n,t,j)}$ by a permutation $\sigma$ of the set $(1,...,n)$ 
and a multi-index ${\bf i}=(i_1,...,i_{n-4})$, where each component takes values in the range $1 \le i_j \le n$ for $1 \le j \le (n-4)$.
We consider terms of the form
\bq
 O^{\mu_1...\mu_n}\left(\sigma,{\bf i}\right)
 = 
 \left( \prod\limits_{k=1}^{j} g^{\mu_{\sigma(2k-1)}\mu_{\sigma(2k)}} \right)
 \left( \prod\limits_{k=1}^{n-2j} \partial_{i_k}^{\mu_{\sigma(2j+k)}} \right)
 \prod\limits_{k=1}^{j-2} \left( \partial_{i_{n-2j+2k-1}} \cdot \partial_{i_{n-2j+2k}} \right).
\eq
For each term of this form we generate the Feynman rule. The procedure how to obtain the Feynman rule 
is outlined in appendix~\ref{appendix:colour_ordered_rules}.
We consider two terms to be equivalent if they lead in the on-shell kinematics and up to a sign to the same Feynman rule.
For each equivalence class we keep only one representative.
At this point we make a choice.
Let us define a function $\theta(\sigma,{\bf i})$, which takes the values one and zero, depending whether the corresponding term is kept or not.
We therefore have the following ansatz for $O^{\mu_1...\mu_n}_{(n,t,j)}$
\bq
 O^{\mu_1...\mu_n}_{(n,t,j)}
 & = &
 \sum\limits_{\sigma} \sum\limits_{\bf i}
 c_{n,t,j,\sigma,{\bf i}} \theta(\sigma,{\bf i}) O^{\mu_1...\mu_n}\left(\sigma,{\bf i}\right),
\eq
with unknown coefficients $c_{n,t,j,\sigma,{\bf i}}$.
\item We insert this ansatz into the $t_{\mathrm{max}}$ BCJ-relations defined by the set in eq.~(\ref{set_topologies}).
We extract the coefficients of the monomials in the independent scalar products of $\eps_i \cdot \eps_j$, $\eps_i \cdot p_j$ and
$p_i \cdot p_j$.
Requiring that these coefficients vanish defines a system of linear equations for the unknown coefficients $c_{n,t,j,\sigma,{\bf i}}$.
We then solve for the coefficients $c_{n,t,j,\sigma,{\bf i}}$.
Usually the solution will not be unique. 
For $n=5$ this non-uniqueness corresponds to the freedom of adding terms of the form as in eq.~(\ref{freedom_five}).
We are only interested in one specific solution.
We therefore make a choice and pick one solution.
This defines $O^{\mu_1...\mu_n}_{(n,t,j)}$ and in turn ${\cal L}^{(n)}$.
\end{enumerate}
We remark that in step 2 not all possible combinations of $\sigma$ and ${\bf i}$ have to be considered.
Due to permutation symmetries and momentum conservation we can impose the following restrictions:
\bq
\begin{array}{llll}
 \sigma(2k-1) < \sigma(2k), & k \in \{1,...,j\}, & \mbox{since} & g^{\mu_{\sigma(2k-1)}\mu_{\sigma(2k)}}=g^{\mu_{\sigma(2k)}\mu_{\sigma(2k-1)}}, \\
 \sigma(2k-1) < \sigma(2k+1), & k \in \{1,...,j-1\}, & \mbox{since} & 
 g^{\mu_{\sigma(2k-1)}\mu_{\sigma(2k)}} g^{\mu_{\sigma(2k+1)}\mu_{\sigma(2k+2)}} \\
 & & & = g^{\mu_{\sigma(2k+1)}\mu_{\sigma(2k+2)}} g^{\mu_{\sigma(2k-1)}\mu_{\sigma(2k)}}, \\
 \sigma(2j+k) < \sigma(2j+k+1), & k \in \{1,..., n-2j \}, & \mbox{since} & 
 \partial_{i_k}^{\mu_{\sigma(2j+k)}} \partial_{i_{k+1}}^{\mu_{\sigma(2j+k+1)}} = \partial_{i_{k+1}}^{\mu_{\sigma(2j+k+1)}} \partial_{i_k}^{\mu_{\sigma(2j+k)}}, \\
 i_{n-2j+2k-1} < i_{n-2j+2k} & k \in \{1,...,j-2\} & \mbox{since} & \partial_{i_{l-1}} \cdot \partial_{i_{l}} = \partial_{i_{l}} \cdot \partial_{i_{l-1}}, \\
 i_{n-2j+2k-1} < i_{n-2j+2k+1} & k \in \{1,...,j-3\} & \mbox{since} & \left( \partial_{i_{l-1}} \cdot \partial_{i_{l}} \right) \left( \partial_{i_{l+1}} \cdot \partial_{i_{l+2}} \right) \\
 & & & = \left( \partial_{i_{l+1}} \cdot \partial_{i_{l+2}} \right) \left( \partial_{i_{l-1}} \cdot \partial_{i_{l}} \right), \\
 i_k < n & k \in \{1, ..., n-4\} & & \mbox{momentum conservation}.
\end{array}
 \nonumber 
\eq
These restrictions are helpful in speeding up the computation.

We further remark that the system of linear equations in step 3 has a block triangular form.
We can first consider only monomials involving exactly two scalar products of the type $\eps_{i_1} \cdot \eps_{i_2}$.
The coefficients of these monomials will yield equations involving only the variables $c_{n,t,2,\sigma,{\bf i}}$.
In general, the coefficient of a monomial involving $j$ scalar products of the type $\eps_{i_1} \cdot \eps_{i_2}$ will
yield equations with variables $c_{n,t,j',\sigma,{\bf i}}$ with $j' \le j$.

\subsection{Results}

We now present an effective Lagrangian of the form 
\bq
{\cal L}_{\mathrm{YM}} + {\cal L}_{\mathrm{GF}} & = & 
 \frac{1}{2g^2} \sum\limits_{n=2}^\infty {\cal L}^{(n)}
\eq
up to $n=6$.
This Lagrangian generated the correct BCJ relations up to $n=6$.
The terms ${\cal L}^{(2)}$ and ${\cal L}^{(3)}$ read
\bq
 {\cal L}^{(2)} = 
 -2 \mbox{Tr} \; {\bf A}_{\mu} \Box {\bf A}^{\mu},
 & &
 {\cal L}^{(3)} = 
 4 \mbox{Tr} \; \left( \partial_\mu {\bf A}_{\nu} \right) \left[ {\bf A}^{\mu}, {\bf A}^{\nu} \right].
\eq
The term ${\cal L}^{(4)}$ has been given in eq.~(\ref{modify_L4_v2})
\bq
 {\cal L}^{(4)} & = &
 O^{\mu_1\mu_2\mu_3\mu_4}
 D^{-1}
 \; \mbox{Tr} \; \left[ {\bf A}_{\mu_1}, {\bf A}_{\mu_2} \right] \left[ {\bf A}_{\mu_3}, {\bf A}_{\mu_4} \right],
 \nonumber \\
O^{\mu_1\mu_2\mu_3\mu_4} & = &
 g^{\mu_1\mu_3} g^{\mu_2\mu_4} g_{\nu_1\nu_2} \partial^{\nu_1}_{12} \partial^{\nu_2}_{34}
\eq
The terms ${\cal L}^{(5)}$ and ${\cal L}^{(6)}$ are of the form
\bq
 {\cal L}^{(n)} & = &
 \sum\limits_t \sum\limits_{j=2}^{[n/2]}
 O^{\mu_1...\mu_n}_{(n,t,j)}
 \;
 \hat{D}^{-1}
 \;
 \mbox{Tr} \; {\bf T}_{\mu_1...\mu_n}^{(n,t)}.
\eq
For $n=5$ there is only one tree structure and also the sum over $j$ reduces to a single term. 
Thus $t=1$ and $j=2$. 
The tree structure is given by
\bq
 \mbox{Tr} \; {\bf T}_{\mu_1\mu_2\mu_3\mu_4\mu_5}^{(5,1)}
 & = &
  \mbox{Tr} \; J \left( \left[ {\bf A}_{\mu_1}, {\bf A}_{\mu_2} \right], {\bf A}_{\mu_3}, {\bf A}_{\mu_4} \right) {\bf A}_{\mu_5}.
\eq
The operator $O^{\mu_1\mu_2\mu_3\mu_4\mu_5}_{(5,1,2)}$ is not unique and we make the choice
already given in eq.~(\ref{choice_5}):
\bq
 O^{\mu_1\mu_2\mu_3\mu_4\mu_5}_{(5,1,2)}
  & = &
  -4 g^{\mu_1\mu_3} g^{\mu_2\mu_4} \partial_{1}^{\mu_5},
\eq
For $n=6$ we have two tree structures
\bq
 \mbox{Tr} \; {\bf T}_{\mu_1\mu_2\mu_3\mu_4\mu_5\mu_6}^{(6,1)}
  & = &
  \mbox{Tr} \; J \left( \left[ \left[ {\bf A}_{\mu_1}, {\bf A}_{\mu_2} \right], {\bf A}_{\mu_3} \right], {\bf A}_{\mu_4}, {\bf A}_{\mu_5} \right) {\bf A}_{\mu_6},
 \nonumber \\
 \mbox{Tr} \; {\bf T}_{\mu_1\mu_2\mu_3\mu_4\mu_5\mu_6}^{(6,2)}
  & = &
  \mbox{Tr} \; J \left( \left[ {\bf A}_{\mu_1}, {\bf A}_{\mu_2} \right], \left[ {\bf A}_{\mu_3}, {\bf A}_{\mu_4} \right], {\bf A}_{\mu_5} \right) {\bf A}_{\mu_6}.
\eq
In addition we can now have $j=2$ or $j=3$.
We thus have four operators $O^{\mu_1\mu_2\mu_3\mu_4\mu_5\mu_6}_{(6,t,j)}$ with $t\in\{1,2\}$ and $j\in\{2,3\}$.
A possible choice is
\bq
\lefteqn{
 O^{\mu_1\mu_2\mu_3\mu_4\mu_5\mu_6}_{(6,1,2)} = } & &
 \nonumber \\
 &&
  - 8 g^{\mu_{1} \mu_{2}} g^{\mu_{3} \mu_{4}} \partial_{1}^{\mu_{5}} \partial_{2}^{\mu_{6}}
 - 4 g^{\mu_{1} \mu_{2}} g^{\mu_{3} \mu_{4}} \partial_{1}^{\mu_{5}} \partial_{4}^{\mu_{6}}
 - 4 g^{\mu_{1} \mu_{2}} g^{\mu_{3} \mu_{4}} \partial_{1}^{\mu_{5}} \partial_{5}^{\mu_{6}}
 + 4 g^{\mu_{1} \mu_{2}} g^{\mu_{4} \mu_{5}} \partial_{1}^{\mu_{3}} \partial_{4}^{\mu_{6}}
 \nonumber \\
 &&
 + 8 g^{\mu_{1} \mu_{2}} g^{\mu_{4} \mu_{5}} \partial_{4}^{\mu_{3}} \partial_{1}^{\mu_{6}}
 + 8 g^{\mu_{1} \mu_{3}} g^{\mu_{2} \mu_{4}} \partial_{2}^{\mu_{5}} \partial_{3}^{\mu_{6}}
 - 24 g^{\mu_{1} \mu_{3}} g^{\mu_{2} \mu_{4}} \partial_{2}^{\mu_{5}} \partial_{4}^{\mu_{6}}
 - 8 g^{\mu_{1} \mu_{3}} g^{\mu_{2} \mu_{4}} \partial_{2}^{\mu_{5}} \partial_{5}^{\mu_{6}}
 \nonumber \\
 &&
 + 8 g^{\mu_{1} \mu_{3}} g^{\mu_{2} \mu_{4}} \partial_{4}^{\mu_{5}} \partial_{5}^{\mu_{6}}
 + 8 g^{\mu_{1} \mu_{3}} g^{\mu_{4} \mu_{5}} \partial_{1}^{\mu_{2}} \partial_{4}^{\mu_{6}}
 - 8 g^{\mu_{1} \mu_{3}} g^{\mu_{4} \mu_{5}} \partial_{4}^{\mu_{2}} \partial_{2}^{\mu_{6}}
 - 8 g^{\mu_{1} \mu_{3}} g^{\mu_{4} \mu_{5}} \partial_{4}^{\mu_{2}} \partial_{3}^{\mu_{6}}
 \nonumber \\
 &&
 + 8 g^{\mu_{1} \mu_{4}} g^{\mu_{2} \mu_{5}} \partial_{1}^{\mu_{3}} \partial_{4}^{\mu_{6}}
 + 24 g^{\mu_{1} \mu_{4}} g^{\mu_{2} \mu_{5}} \partial_{4}^{\mu_{3}} \partial_{2}^{\mu_{6}}
 + 16 g^{\mu_{1} \mu_{4}} g^{\mu_{2} \mu_{5}} \partial_{4}^{\mu_{3}} \partial_{3}^{\mu_{6}}
 + 32 g^{\mu_{1} \mu_{4}} g^{\mu_{2} \mu_{5}} \partial_{4}^{\mu_{3}} \partial_{4}^{\mu_{6}}
 \nonumber \\
 &&
 + 32 g^{\mu_{1} \mu_{4}} g^{\mu_{2} \mu_{5}} \partial_{4}^{\mu_{3}} \partial_{5}^{\mu_{6}}
 + 4 g^{\mu_{1} \mu_{4}} g^{\mu_{2} \mu_{6}} \partial_{5}^{\mu_{3}} \partial_{3}^{\mu_{5}}
 - 16 g^{\mu_{1} \mu_{4}} g^{\mu_{2} \mu_{6}} \partial_{5}^{\mu_{3}} \partial_{4}^{\mu_{5}}
 - 8 g^{\mu_{1} \mu_{4}} g^{\mu_{3} \mu_{5}} \partial_{1}^{\mu_{2}} \partial_{5}^{\mu_{6}}
 \nonumber \\
 &&
 + 8 g^{\mu_{1} \mu_{4}} g^{\mu_{3} \mu_{5}} \partial_{3}^{\mu_{2}} \partial_{1}^{\mu_{6}}
 + 8 g^{\mu_{1} \mu_{4}} g^{\mu_{3} \mu_{5}} \partial_{3}^{\mu_{2}} \partial_{2}^{\mu_{6}}
 + 8 g^{\mu_{1} \mu_{4}} g^{\mu_{3} \mu_{5}} \partial_{3}^{\mu_{2}} \partial_{3}^{\mu_{6}}
 + 16 g^{\mu_{1} \mu_{4}} g^{\mu_{3} \mu_{5}} \partial_{5}^{\mu_{2}} \partial_{1}^{\mu_{6}}
 \nonumber \\
 &&
 + 8 g^{\mu_{1} \mu_{4}} g^{\mu_{3} \mu_{5}} \partial_{5}^{\mu_{2}} \partial_{2}^{\mu_{6}}
 - 8 g^{\mu_{1} \mu_{4}} g^{\mu_{3} \mu_{5}} \partial_{5}^{\mu_{2}} \partial_{3}^{\mu_{6}}
 - 8 g^{\mu_{1} \mu_{4}} g^{\mu_{3} \mu_{6}} \partial_{5}^{\mu_{2}} \partial_{1}^{\mu_{5}}
 + 16 g^{\mu_{1} \mu_{4}} g^{\mu_{5} \mu_{6}} \partial_{5}^{\mu_{2}} \partial_{1}^{\mu_{3}}
 \nonumber \\
 &&
 - 16 g^{\mu_{3} \mu_{4}} g^{\mu_{5} \mu_{6}} \partial_{2}^{\mu_{1}} \partial_{5}^{\mu_{2}}
 - 16 g^{\mu_{3} \mu_{4}} g^{\mu_{5} \mu_{6}} \partial_{4}^{\mu_{1}} \partial_{5}^{\mu_{2}}
 + 16 g^{\mu_{3} \mu_{6}} g^{\mu_{4} \mu_{5}} \partial_{4}^{\mu_{1}} \partial_{5}^{\mu_{2}},
 \nonumber \\
\lefteqn{
 O^{\mu_1\mu_2\mu_3\mu_4\mu_5\mu_6}_{(6,2,2)} = } & &
 \nonumber \\
 &&
 - 2 g^{\mu_{1} \mu_{2}} g^{\mu_{3} \mu_{4}} \partial_{1}^{\mu_{5}} \partial_{3}^{\mu_{6}}
 - 12 g^{\mu_{1} \mu_{2}} g^{\mu_{3} \mu_{5}} \partial_{1}^{\mu_{4}} \partial_{3}^{\mu_{6}}
 - 4 g^{\mu_{1} \mu_{2}} g^{\mu_{3} \mu_{5}} \partial_{1}^{\mu_{4}} \partial_{4}^{\mu_{6}}
 - 8 g^{\mu_{1} \mu_{2}} g^{\mu_{5} \mu_{6}} \partial_{1}^{\mu_{3}} \partial_{5}^{\mu_{4}}
 \nonumber \\
 &&
 - 12 g^{\mu_{1} \mu_{3}} g^{\mu_{2} \mu_{4}} \partial_{1}^{\mu_{5}} \partial_{3}^{\mu_{6}}
 - 4 g^{\mu_{1} \mu_{3}} g^{\mu_{2} \mu_{4}} \partial_{1}^{\mu_{5}} \partial_{5}^{\mu_{6}}
 - 8 g^{\mu_{1} \mu_{3}} g^{\mu_{2} \mu_{5}} \partial_{1}^{\mu_{4}} \partial_{4}^{\mu_{6}}
 + 16 g^{\mu_{1} \mu_{3}} g^{\mu_{2} \mu_{5}} \partial_{3}^{\mu_{4}} \partial_{1}^{\mu_{6}}
 \nonumber \\
 &&
 - 8 g^{\mu_{1} \mu_{3}} g^{\mu_{2} \mu_{5}} \partial_{5}^{\mu_{4}} \partial_{4}^{\mu_{6}}
 + 8 g^{\mu_{1} \mu_{3}} g^{\mu_{5} \mu_{6}} \partial_{1}^{\mu_{2}} \partial_{5}^{\mu_{4}}
 + 8 g^{\mu_{1} \mu_{5}} g^{\mu_{2} \mu_{6}} \partial_{1}^{\mu_{3}} \partial_{3}^{\mu_{4}}
 + 8 g^{\mu_{1} \mu_{5}} g^{\mu_{2} \mu_{6}} \partial_{1}^{\mu_{3}} \partial_{5}^{\mu_{4}}
 \nonumber \\
 &&
 - 8 g^{\mu_{1} \mu_{5}} g^{\mu_{3} \mu_{6}} \partial_{1}^{\mu_{2}} \partial_{2}^{\mu_{4}}
 - 4 g^{\mu_{1} \mu_{5}} g^{\mu_{3} \mu_{6}} \partial_{1}^{\mu_{2}} \partial_{3}^{\mu_{4}}
 - 16 g^{\mu_{1} \mu_{5}} g^{\mu_{3} \mu_{6}} \partial_{1}^{\mu_{2}} \partial_{5}^{\mu_{4}}
 + 12 g^{\mu_{1} \mu_{5}} g^{\mu_{3} \mu_{6}} \partial_{3}^{\mu_{2}} \partial_{1}^{\mu_{4}}
 \nonumber \\
 &&
 + 8 g^{\mu_{1} \mu_{5}} g^{\mu_{3} \mu_{6}} \partial_{3}^{\mu_{2}} \partial_{2}^{\mu_{4}}
 - 4 g^{\mu_{1} \mu_{5}} g^{\mu_{3} \mu_{6}} \partial_{4}^{\mu_{2}} \partial_{2}^{\mu_{4}}
 - 8 g^{\mu_{1} \mu_{5}} g^{\mu_{3} \mu_{6}} \partial_{5}^{\mu_{2}} \partial_{3}^{\mu_{4}},
 \nonumber \\
\lefteqn{
 O^{\mu_1\mu_2\mu_3\mu_4\mu_5\mu_6}_{(6,1,3)} = } & &
 \nonumber \\
 &&
 - 2 g^{\mu_{1} \mu_{4}} g^{\mu_{2} \mu_{5}} g^{\mu_{3} \mu_{6}} \left( \partial_{1} \cdot \partial_{4} \right)
 - 4 g^{\mu_{1} \mu_{4}} g^{\mu_{2} \mu_{5}} g^{\mu_{3} \mu_{6}} \left( \partial_{1} \cdot \partial_{5} \right)
 - 6 g^{\mu_{1} \mu_{4}} g^{\mu_{2} \mu_{5}} g^{\mu_{3} \mu_{6}} \left( \partial_{4} \cdot \partial_{5} \right),
 \nonumber \\
\lefteqn{
 O^{\mu_1\mu_2\mu_3\mu_4\mu_5\mu_6}_{(6,2,3)} = } & &
 \nonumber \\
 &&
 - 2 g^{\mu_{1} \mu_{3}} g^{\mu_{2} \mu_{4}} g^{\mu_{5} \mu_{6}} \left( \partial_{1} \cdot \partial_{5} \right)
 - 2 g^{\mu_{1} \mu_{3}} g^{\mu_{2} \mu_{5}} g^{\mu_{4} \mu_{6}} \left( \partial_{1} \cdot \partial_{3} \right)
 - 2 g^{\mu_{1} \mu_{3}} g^{\mu_{2} \mu_{5}} g^{\mu_{4} \mu_{6}} \left( \partial_{2} \cdot \partial_{4} \right)
 \nonumber \\
 &&
 + 2 g^{\mu_{1} \mu_{3}} g^{\mu_{2} \mu_{5}} g^{\mu_{4} \mu_{6}} \left( \partial_{2} \cdot \partial_{5} \right)
 + 4 g^{\mu_{1} \mu_{3}} g^{\mu_{2} \mu_{5}} g^{\mu_{4} \mu_{6}} \left( \partial_{3} \cdot \partial_{5} \right).
\eq
In addition we obtain the full set of additional operators at $n=6$, which can be added to the Lagrangian without violating the 
BCJ relations. 
These operators are the analog of the five-point operator given in eq.~(\ref{freedom_five}).
This list, as well as all results for $n>6$ are rather lengthy and not presented here.

\section{Conclusions}
\label{sect:conclusions}

In this paper we have presented a systematic method to construct an effective Lagrangian which 
(i) agrees with the standard Yang-Mills Lagrangian in the sense that the difference between the two Lagrangians simplifies to zero,
and which
(ii) generates the BCJ decomposition automatically.
The second property is the non-trivial property of our Lagrangian.
The algorithm presented here will be useful for a more detailed study of the BCJ relations and in particular for the implications for gravity.
We presented explicit results for $n$ up to $6$.

\subsection*{Acknowledgements}

M.T. acknowledges support through a fellowship from the GRK {\it Symmetry Breaking} (DFG/GRK 1581).
S.W. would like to thank the Kavli Institute for Theoretical Physics for hospitality, where this project was initiated.
This research was supported in part by the National Science Foundation under Grant No. NSF PHY05-51164.


\begin{appendix}


\section{Feynman rules}
\label{appendix:colour_ordered_rules}

Suppose the interaction Lagrangian is given by
\bq
 {\cal L}_{\mathrm{int}} & = & 
 O^{a_1...a_n,\mu_1...\mu_n}\left(\partial_1,...,\partial_n \right)
 A_{\mu_1}^{a_1}(x) ... A_{\mu_n}^{a_n}(x),
\eq
where $O^{a_1...a_n,\mu_1...\mu_n}$ is a pseudo-differential operator of degree $r=(4-n)$ and
$\partial_j$ is a derivative acting only on the field $A_{\mu_j}^{a_j}$. 
Then the full Feynman rule for the vertex (including colour) is
\bq
 V_{\mathrm{full}} & = &
 i \sum\limits_{\sigma \in S_n} O^{a_{\sigma(1)}...a_{\sigma(n)},\mu_{\sigma(1)}...\mu_{\sigma(n)}}\left(i p_{\sigma(1)}, ..., i p_{\sigma(n)}\right),
\eq
where the momenta $p_j$ are out-going and the sum is over all permutations.
Suppose now that
\bq
 O^{a_1...a_n,\mu_1...\mu_n}\left(\partial_1,...,\partial_n \right)
 & = &
 2 g^{n-2} \;\mbox{Tr}\left(T^{a_1} ... T^{a_n}\right)
 O^{\mu_1...\mu_n}\left(\partial_1,...,\partial_n \right).
\eq
We can split the summation over $S_n$ into $S_n/{\mathbb Z}_n$ and ${\mathbb Z}_n$ and obtain
\bq
 V_{\mathrm{full}} & = &
 i 
 g^{n-2}
 \sum\limits_{\pi \in S_n/{\mathbb Z}_n} 
 2 \;\mbox{Tr}\left(T^{a_{\pi(1)}} ... T^{a_{\pi(n)}}\right)
 \sum\limits_{\sigma \in {\mathbb Z}_n} 
 O^{\mu_{\sigma \pi(1)}...\mu_{\sigma \pi(n)}}\left(i p_{\sigma \pi(1)}, ..., i p_{\sigma \pi(n)}\right).
\eq
The colour-ordered Feynman rule reads
\bq
 V & = &
 i 
 \sum\limits_{\sigma \in {\mathbb Z}_n} 
 O^{\mu_{\sigma(1)}...\mu_{\sigma(n)}}\left(i p_{\sigma(1)}, ..., i p_{\sigma(n)}\right).
\eq
Let us now consider a tree $T$ with $n$ external legs in the cyclic order $(1,...,n)$ 
and let us denote by ${\bf T}_{\mu_1...\mu_n}$ a representation of this tree made out of commutators and fields ${\bf A}_{\mu_i}$.
To give an example the representation of the tree
\bq
 T & = & \left( \left[\left[1,2\right],\left[3,4\right]\right], 5 \right)
\eq
is given by
\bq
 {\bf T}_{\mu_1\mu_2\mu_3\mu_4\mu_5} & = &
 \left[ \left[ {\bf A}_{\mu_1}, {\bf A}_{\mu_2} \right], \left[ {\bf A}_{\mu_3}, {\bf A}_{\mu_4} \right] \right] {\bf A}_{\mu_5}.
\eq
The tree $T$ has $(n-2)$ vertices and when drawn as a rooted tree there are $2^{n-2}$ ways of swapping at each vertex the two branches not connected to the root.
Each possibility defines a different cyclic order of the external legs.
Hence we obtain a set of permutations of the external legs, which we denote by $B_n(T)$.
Let us now consider a term of the form
\bq
{\cal L}_{\mathrm{int}}
 & = &
 \frac{1}{2g^2} 
 O^{\mu_1...\mu_n}\left(\partial_1,...,\partial_n\right)
 \;\mbox{Tr} \; {\bf T}_{\mu_1...\mu_n}.
\eq
Using the anti-symmetry of the commutators we can write this term equally well as
\bq
{\cal L}_{\mathrm{int}}
 & = &
 \frac{1}{2g^2} 
 O^{\mu_1...\mu_n}\left(\partial_1,...,\partial_n\right)
 \sum\limits_{\pi \in B_n(T)} \left(-1\right)^{n_{\mathrm{swap}}(\pi)}
 \;\mbox{Tr} \; {\bf A}_{\mu_{\pi(1)}} ... {\bf A}_{\mu_{\pi(n)}},
\eq
where $n_{\mathrm{swap}}(\pi)$ denotes the number of swaps needed to arrive at the permutation $\pi$.
To give an example consider the tree $([[1,2],3],4)$. Swapping $[1,2]$ with $3$ leads to $([3,[1,2]],4)$ and $n_{\mathrm{swap}}=1$.
Hence, $(-1)^{n_{\mathrm{swap}}}=-1$. This should not be confused with the sign of the permutation $\pi=(3,1,2,4)$, which is $(+1)$.

Relabelling the indices we arrive at
\bq
{\cal L}_{\mathrm{int}}
 & = &
 \frac{1}{2g^2} 
 \sum\limits_{\pi \in B_n(T)} \left(-1\right)^{n_{\mathrm{swap}}(\pi)}
 O^{\mu_{\pi^{-1}(1)}...\mu_{\pi^{-1}(n)}}\left(\partial_{\pi^{-1}(1)},...,\partial_{\pi^{-1}(n)}\right)
 \;\mbox{Tr} \; {\bf A}_{\mu_1} ... {\bf A}_{\mu_n}
\eq
and at the colour-ordered Feynman rule
\bq
 V & = &
 \frac{(-1)^n}{4} i^{n+1}
 \sum\limits_{\pi \in B_n(T)} \left(-1\right)^{n_{\mathrm{swap}}(\pi)}
 \sum\limits_{\sigma \in {\mathbb Z}_n} 
 O^{\mu_{\sigma\pi^{-1}(1)}...\mu_{\sigma\pi^{-1}(n)}}\left(i p_{\sigma\pi^{-1}(1)},...,i p_{\sigma\pi^{-1}(n)}\right).
\eq
Next we consider an interaction Lagrangian of the form
\bq
\label{bcj_vertex}
{\cal L}_{\mathrm{int}}
 & = &
 \frac{1}{2g^2} 
 O^{\mu_1...\mu_n}\left(\partial_1,...,\partial_n\right)
 \;
 \mbox{Tr} \; {\bf T}_{\mu_1...\mu_n},
\eq
with
\bq
 {\bf T}_{\mu_1...\mu_n} & = & 
 J\left( {\bf T}_{\mu_1...\mu_{j_1}}, {\bf T}_{\mu_{j_1+1}...\mu_{j_2}}, {\bf T}_{\mu_{j_2+1}...\mu_{j_3}} \right) {\bf T}_{\mu_{j_3+1}...\mu_n}
\eq
The symbol $J$ denotes the sum over the three permutations of the Jacobi identity.
It is convenient to define three trees, corresponding to the three terms in the Jacobi identity. We denote these trees by $T_{1234}$, $T_{2314}$ and $T_{3124}$.
They correspond to
\bq
 T_{1234} & : & \left[ \left[ {\bf T}_{\mu_1...\mu_{j_1}}, {\bf T}_{\mu_{j_1+1}...\mu_{j_2}} \right], {\bf T}_{\mu_{j_2+1}...\mu_{j_3}}\right] {\bf T}_{\mu_{j_3+1}...\mu_n}
 \nonumber \\
 T_{2314} & : & \left[ \left[ {\bf T}_{\mu_{j_1+1}...\mu_{j_2}} , {\bf T}_{\mu_{j_2+1}...\mu_{j_3}} \right], {\bf T}_{\mu_1...\mu_{j_1}} \right] {\bf T}_{\mu_{j_3+1}...\mu_n}
 \nonumber \\
 T_{3124} & : & \left[ \left[ {\bf T}_{\mu_{j_2+1}...\mu_{j_3}}, {\bf T}_{\mu_1...\mu_{j_1}} \right], {\bf T}_{\mu_{j_1+1}...\mu_{j_2}} \right] {\bf T}_{\mu_{j_3+1}...\mu_n}
\eq
In addition we define three permutations by
\bq
 \tau_{1234} & = & \left( 1, ..., n \right),
 \nonumber \\
 \tau_{2314} & = & \left( j_1+1,...,j_2,j_2+1,...,j_3,1,...,j_1,j_3+1,...,n \right),
 \nonumber \\
 \tau_{3124} & = & \left( j_2+1,...j_3, 1,..., j_1, j_1+1,...,j_2, j_3+1, ..., n \right).
\eq
We further denote 
\bq
 I & = & \left\{ 1234, 2314, 3124 \right\}.
\eq
We need a little bit more of notation.
We denote by $T(1,...,n)$ a tree with the cyclic order $(1,...,n)$.
By abuse of notation we denote by $(\pi T)$
the tree we obtain from the original tree $T$ by performing all swaps to arrive at the permutation $\pi$. $(\pi T)$ has then the cyclic order
$(\pi(1),...,\pi(n))$. We further denote by
\bq
 \left( \pi T \right)\left(1,...,n\right)
\eq
the above constructed tree $(\pi T)$, where the external legs have been relabelled as $(1,...,n)$.
Similarly, we denote by $(\tau_j T)$ ($j \in I$) the tree obtained from $T$ by performing the Jacobi operation corresponding to $\tau_j$.
This tree has then the cyclic order $(\tau(1),...,\tau(n))$. 
With the definitions as above we have for example $T_{2314} = \tau_{2314} T_{1234}$.
We further denote by
\bq
 \left( \tau T \right)\left(1,...,n\right)
\eq
the tree $(\tau T)$, where the external legs have been relabelled as $(1,...,n)$.
The colour-ordered Feynman rule for the vertex in eq.~(\ref{bcj_vertex}) is with $T=T_{1234}$
\bq
\label{rule_bcj}
 V & = &
 \frac{(-1)^n}{4} i^{n+1}
 \sum\limits_{j \in I}
 \;\;\;
 \sum\limits_{\pi \in B_n((\tau_j T)(1,..,n))} \left(-1\right)^{n_{\mathrm{swap}}(\pi)}
 \\
 & &
 \sum\limits_{\sigma \in {\mathbb Z}_n} 
 O^{\mu_{\sigma\pi^{-1}\tau_j^{-1}(1)}...\mu_{\sigma\pi^{-1}\tau_j^{-1}(n)}}\left(i p_{\sigma\pi^{-1}\tau_j^{-1}(1)},...,i p_{\sigma\pi^{-1}\tau_j^{-1}(n)}\right).
\eq
Finally, we consider an interaction term in the Lagrangian of the form
\bq
\label{bcj_vertex2}
{\cal L}_{\mathrm{int}}
 & = &
 \frac{1}{2g^2} 
 O^{\mu_1...\mu_n}\left(\partial_1,...,\partial_n\right)
 \;
 \hat{D}^{-1}
 \;
 \mbox{Tr} \; {\bf T}_{\mu_1...\mu_n},
\eq
Eq.~(\ref{bcj_vertex2}) differs from eq.~(\ref{bcj_vertex}) only through the explicit appearance of the operator $\hat{D}^{-1}$.
In principle, we could think of the operator $\hat{D}^{-1}$ as being part of $O^{\mu_1...\mu_n}(\partial_1,...,\partial_n)$ in eq.~(\ref{bcj_vertex}),
resulting in the Feynman rule of eq.~(\ref{rule_bcj}).
However for our purpose it is convenient to show this operator explicitly.
$\hat{D}^{-1}$ depends only on the tree structure it acts on. 
The Feynman rule is then
\bq
 V & = &
 \frac{(-1)^n}{4} i^{n+1}
 \sum\limits_{j \in I}
 \;\;\;
 \sum\limits_{\pi \in B_n((\tau_j T)(1,..,n))} \left(-1\right)^{n_{\mathrm{swap}}(\pi)}
 \\
 & &
 \sum\limits_{\sigma \in {\mathbb Z}_n} 
 O^{\mu_{\sigma\pi^{-1}\tau_j^{-1}(1)}...\mu_{\sigma\pi^{-1}\tau_j^{-1}(n)}}\left(i p_{\sigma\pi^{-1}\tau_j^{-1}(1)},...,i p_{\sigma\pi^{-1}\tau_j^{-1}(n)}\right)
 \hat{D}^{-1}\left(\left(\pi \tau_j T \right)\left(\sigma(1),...,\sigma(n)\right)\right).
 \nonumber 
\eq
In the case where $O$ is a differential operator of degree $(n-4)$ the Feynman rule simplifies to
\bq
 V & = &
 \frac{i}{4} 
 \sum\limits_{j \in I}
 \;\;\;
 \sum\limits_{\pi \in B_n((\tau_j T)(1,..,n))} \left(-1\right)^{n_{\mathrm{swap}}(\pi)}
 \\
 & &
 \sum\limits_{\sigma \in {\mathbb Z}_n} 
 O^{\mu_{\sigma\pi^{-1}\tau_j^{-1}(1)}...\mu_{\sigma\pi^{-1}\tau_j^{-1}(n)}}\left(p_{\sigma\pi^{-1}\tau_j^{-1}(1)},...,p_{\sigma\pi^{-1}\tau_j^{-1}(n)}\right)
 \hat{D}^{-1}\left(\left(\pi \tau_j T \right)\left(\sigma(1),...,\sigma(n)\right)\right).
 \nonumber 
\eq

\section{Inequivalent tree topologies}
\label{appendix:inequivalent_tree_topologies}

In this appendix we discuss how to find the inequivalent tree topologies which we have to consider for the Jacobi relations.
It is clear that for a Jacobi relation we have to consider trees with at least four external legs.
Furthermore, each of the three terms in the Jacobi relation has a marked propagator.
Therefore we can represent one term in the Jacobi relation by
\bq
\label{basic_jacobi_cut_tree}
 T = \left( T_{12}, T_{34} \right) = \left( \left[T_1,T_2\right],\left[T_3,T_4\right]\right).
\eq
The marked propagator corresponds to the root of the sub-tree $T_{12}$ (or equivalently to the root of the sub-tree $T_{34}$).
We call two trees $T$ and $T'$ of the form as in eq.~(\ref{basic_jacobi_cut_tree}) equivalent, if they differ only through the following three operations:
\begin{enumerate}
\item Cyclic property of $(...,...)$:
\bq
 \left( T_{12}, T_{34} \right) & = & \left( T_{34}', T_{12}' \right).
\eq
\item Anti-symmetry of $[...,...]$: $T_{12}'$ ($T_{34}'$) can be obtained from $T_{12}$ ($T_{34}$) by a sequence of swaps of the form
\bq
 \left[ T_a, T_b \right] & \rightarrow & \left[ T_b, T_a \right],
\eq
where $T_a$ and $T_b$ are sub-trees at one vertex of $T_{12}$ ($T_{34}$).
The sign is not important as long as we are only concerned with (in-)equivalent trees.
\item Jacobi operation:
\bq
 \left( \left[T_1,T_2\right],\left[T_3,T_4\right]\right) & = & \left( \left[T_2',T_3'\right],\left[T_1',T_4'\right]\right).
\eq
\end{enumerate}
It is then a relatively easy exercise to write a computer program, which determines all classes of inequivalent trees 
of the form~(\ref{basic_jacobi_cut_tree}).
Within a given class we can choose any member as a representative.
Up to $n=8$ we have the following classes of inequivalent trees:
\bq
n = 4 & : & 
([1,2], [3,4]),
 \nonumber \\
n = 5 & : & 
([[1,2],3], [4,5]),
 \nonumber \\
n = 6 & : & 
([[[1,2],3],4], [5,6]),
 \;\;\;
([[1,2],[3,4]], [5,6]),
 \nonumber \\
n = 7 & : &
([[[[1,2],3],4],5], [6,7]),
 \;\;\;
([[[1,2],[3,4]],5], [6,7]),
 \;\;\;
([[[1,2],3],[4,5]], [6,7]),
 \nonumber \\
 & & 
([[1,2],[3,4]], [[5,6],7]),
 \nonumber \\
n = 8 & : &
([[[[[1,2],3],4],5],6], [7,8]),
 \;\;\;
([[[[1,2],[3,4]],5],6], [7,8]),
 \;\;\;
([[[[1,2],3],[4,5]],6], [7,8]),
 \nonumber \\
 & & 
([[[[1,2],3],4],[5,6]], [7,8]),
 \;\;\;
([[[1,2],[3,4]],[5,6]], [7,8]),
 \;\;\;
([[[1,2],3],[[4,5],6]], [7,8]),
 \nonumber \\
 & & 
([[[1,2],3],[4,5]], [[6,7],8]),
 \;\;\;
([[1,2],[3,4]], [[5,6],[7,8]]).
\eq

\end{appendix}

\bibliography{/home/stefanw/notes/biblio}
\bibliographystyle{/home/stefanw/latex-style/h-physrev5}

\end{document}